# Fermi Level Modulation of Boron Nitride Nanosheets by Vacancy Driven Compressive Strain.

Tanmay Mahanta[1], Tanuja Mohanty[1),a]

*School of Physical Sciences, JNU, New Delhi-110067, INDIA*

[a)]Corresponding author: tanujajnu@gmail.com

## Abstract

The compressive strain induced work function modulation in boron nitride nanosheets (BNNS) has been studied. The compressive strain arose due to the irradiation of 30 keV Au beam on thin films of BNNS. Before irradiation BNNS was characterized by Raman spectroscopy, UV-Vis spectroscopy, where using Tauc plot the bandgap of chemically synthesized BNNS from commercial hBN powder has been calculated as well as the layer numbers on thin films also quantified using an empirical relation exploiting the full width at half maxima (FWHM) of Raman peak of BNNS. The Scanning Electron Microscopy (SEM) images were also taken where layer like structure in pristine BNNS sample is visible and irradiation-induced changes in terms of irregular dots on BNNS samples is witnessed. The X-Ray Diffraction (XRD) was used to characterize the BNNS samples where peak corresponding to (002) plane is depicted as well as the peak shifting to higher angle which indicates the compressive strain in BNNS is also visible. Finally, the possible explanation of compressive strain formation by Au irradiation on BNNS was discussed using SRIM-2013 MC simulations. We've explained that vacancy induces strain in materials that lower the work function (WF). Scanning Kelvin Probe Microscopy (SKPM) was used to map the work function of the surface of BNNS, and it turned out that WF lowers with increasing dose of ions. The value of strain also calculated using the work function values and the results are qualitatively agreeing with the XRD results.

**Keywords:** Hexagonal boron nitride; Ion beam irradiation; UV-Vis spectroscopy; Work Function; Strain.

## 1. Introduction:

To engineer the properties of materials introduction of strain is one of the finest methods in which without altering the generality, selective modifications of materials could be directed controlling the external parameters [1,2]. It was observed theoretically and experimentally in some materials that strain could alter the band gap [3], lead to indirect to direct band gap transition[4,5,6], shifting of phonon modes[7,8], etc. Optical and electronic properties are also moderated upon the imposition of strain. The effects of uniaxial and biaxial strains are different in different materials[9]. There are several methods of imposing strain on target materials, ion beam technology is one of them, it is preferable over conventional techniques as externally controlling the dose of ions, the energy of ions and angle of incidence we can control the changes inside the materials[10, 11, 12., 13]. There's less possibility of the presence of contaminants in target materials too. Hexagonal boron nitride[14], sharing a similar structure with graphene often leveled as white graphene has a large band gap[14,15] falls in the category of an insulator. Recently it was observed experimentally that hBN possesses high plane stiffness and tensile strength[16]. It's widely used in making hetero structure[18] and as a great substrate for graphene[14]. Apart from that in lubricating industry hBN[19] is widely used. Due to its robustness, it could've been used in making long lasting photoelectric devices, but its high work function [20, 21] doesn't meet the criteria as the work function value should be lower and controllable. Like graphene, hBN could also can be synthesized by direct growth method, mechanical exfoliation, ion intercalation method [15, 20, 22]. The exfoliated ultrathin material possesses the quality of the material but in a low dimension, specifically with lower thickness. Chemically exfoliated BNNS generally contains a few layers of hBN [23], useful to use in a wide range of applications including polymer-based composites [24], piezoelectric devices and biosensors[25, 26]. Herein this work we exfoliate BNNS exploiting chemical exfoliation method and tried to observe the strain-induced modifications in hBN. We've observed that ion beam induced strain could impose compressive strain along (002) direction and lowers its work function value. The Raman spectroscopy method was utilized to characterize the material. Shifting of peaks in the XRD plot was used to quantify the strain imposed in the material due to ion beam irradiation. Finally, SKPM was employed to map the work function modulation in chemically exfoliated a few layers BNNS. The possible explanation of work function modulation was also discussed using the vacancy graph computed by SRIM software. The work function value derived strain is also explored approximately which is in good agreement with XRD results.

## 2. Experimental Section:

Bulk hBN powder, Dimethylformamide (DMF) was purchased from Sigma Aldrich. All chemicals were used as obtained. Before sonication in tip and bath sonicator, the commercial hBN powder was ground for a long time in mortar and pestle to generate defect sites in it which in turn helps to exfoliate the layers[]. The finely ground powder was then mixed with DMF and was put in a tip sonicator for six hours. The suspension was then allowed to rest for 24 hours.

The supernatant was then collected and again sonicated in a bath sonicator. The final BNNS suspension was obtained by collecting the supernatant after centrifuging the collected suspension from the bath sonicator for 30 minutes at 3000 rpm. The milky white solution was full of a few layers of BNNS. Thin films of BNNS were obtained by drop-casting BNNS solution on Si substrate.

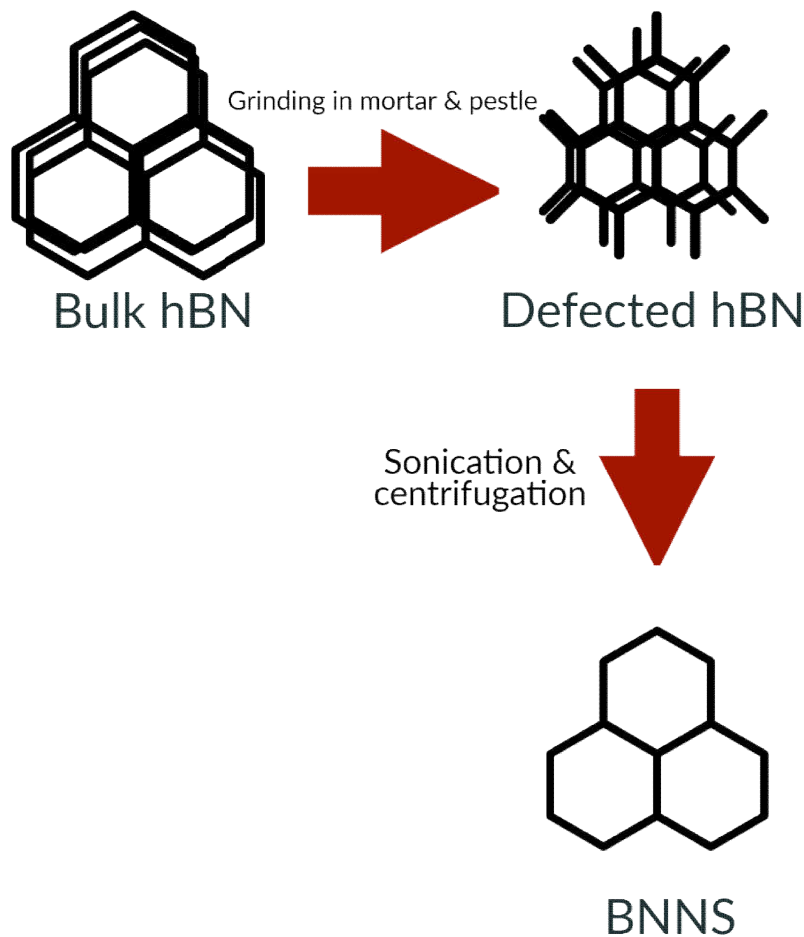

The irradiation of low energy Au beam (30 keV) was done at IUAC, New Delhi (details) PANlytical X'pert PRO, (Cu-Kα line) was used for XRD studies. Raman analysis was done by Witec Alpha 300, (using 532 nm Laser, 100 µW). The UV-VIS spectroscopy was done using Shimadzu UV 2600 UV-VIS Spectrophotometer. The work function was measured in terms of CPD using Scanning Kelvin probe Microscopy (SKPM) of KP Technology. The theoretical calculations were performed using SRIM – 2013 software.

## 3. Results:

### 3.1. Raman spectra:

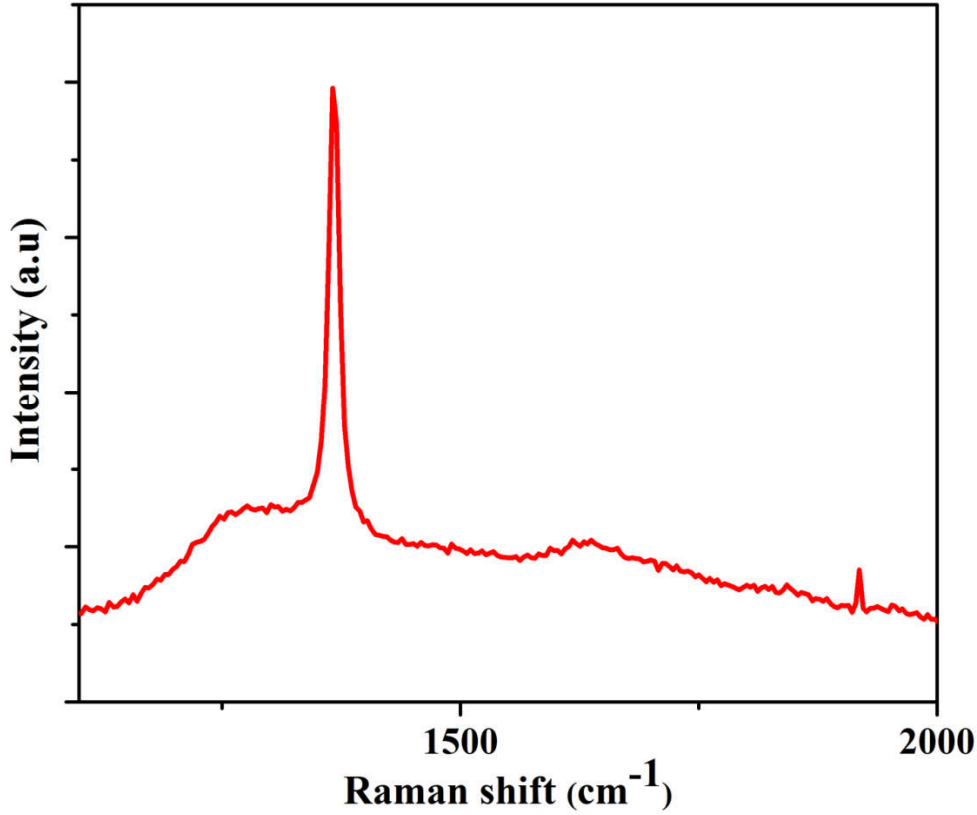

Fig-1: Raman spectra of BNNS. The peak at ~1365 cm$^{-1}$ corresponds to E2g phonon vibration in BNNS.

Raman spectroscopy is a remarkable tool that enables us to identify different material utilizing their characteristic peaks. It is a nondestructive easeful instrumentation system. Raman spectroscopy maps the selective phonon vibrations which are Raman active. Raman spectrum of BNNS thin films had been taken as shown in figure-1. In hBN, the only Raman active mode is at 1365cm$^{-1}$ known as G peak [27]. It's the outcome of $E_{2g}$ vibrations of Phonons in hBN [27]. With the decrement of layer numbers the intensity decreases. hBN is not resonantly excited hence the spectrum is very weak. In a single layer of BNNS, it's almost diminished. In our case the intensity is fairly high, giving rise to a sharp peak at ~1366 cm$^{-1}$. It's a sign that we've succeeded in synthesizing a few layers of BNNS chemically. Using the Raman spectrum, we can find out the number of layers using equation-1 by estimating the FWHM [22].

$$<N>_{vf} = \frac{17.2}{[\Gamma_G - 8.5 - logP]} - 1 \qquad (1)$$

Where, $<N>_{vf}$ is volume fraction weighted mean layer number, which relates to mean layer number $<N>$ with a ratio of 1.5[22]. So, they are linearly related. P is the laser power used. $\Gamma_G$ is the FWHM of Raman G mode of BNNS, which has a value after Lorentzian fitting 12.6 cm$^{-1}$. The laser power was 4 mW. After putting the values in the equation we've found that the thin films consisted of ~4 layers of BNNS.

## 3.2. SEM images:

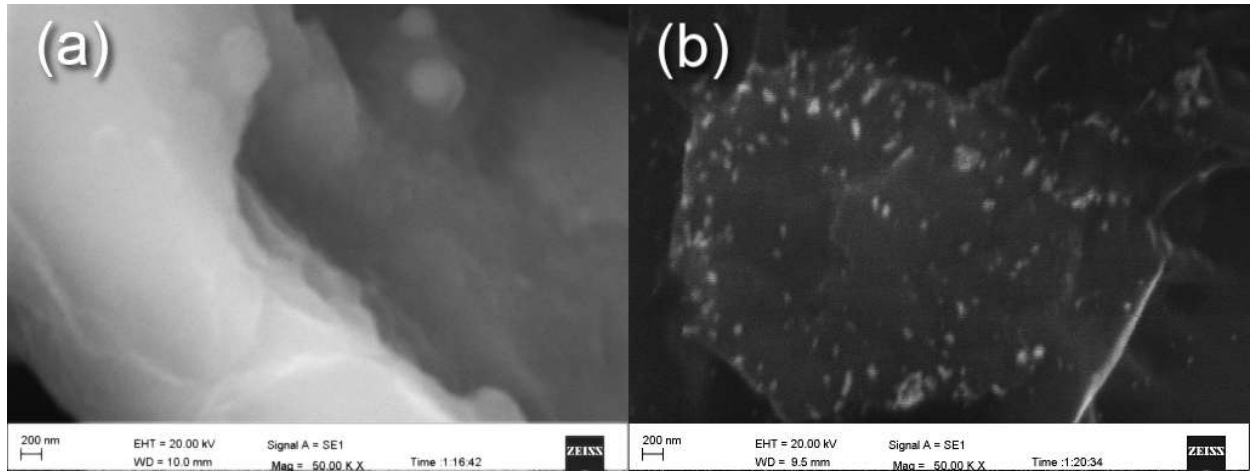

Fig-2: SEM images of (a) pristine and (b) irradiated BNNS (1E16).

The SEM images of the pristine BNNS sample and one irradiated BNNS sample were taken. Slightly visible yet the presence of BNNS on the substrate can be confirmed. The amalgamation of BNNS may be due to connected with drop cast method of thin-film production. The irradiated BNNS can be confirmed to be damaged on its upper surface as visible from the image. The Au ion produced damages is noticeable clearly as irregular spots on BNNS surfaces.

## 3.3. UV-Vis spectra:

The UV-Vis spectroscopy is another remarkable technique which is quite helpful in identifying materials using light of wavelength varying from ultraviolet to visible region. When the light of a wavelength from visible to ultraviolet is allowed to pass through any material depending upon the orbital structure of the materials, light of a certain wavelength is absorbed which reflects as a peak in UV-Vis spectra. Therefore UV-Vis spectroscopy is useful to determine the optical band gap also. hBN, being a member of the insulator family, has a direct bandgap in the ultraviolet range. The bandgap of hBN depends upon on the layer number. For bulk hBN, it's ranging from 4 to 5 eV, whereas with the decrease of layer number the value increases. In fig 3, the UV-Vis spectrum of as-synthesized BNNS is depicted.

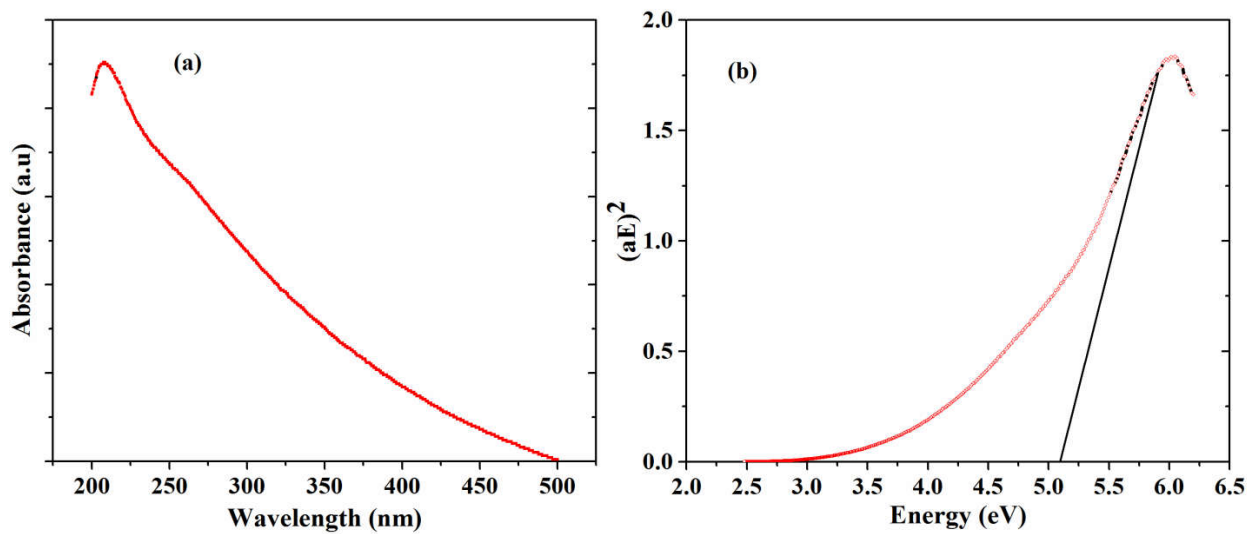

Fig-3: (a) UV-Vis spectrum of synthesized BNNS. (b) Tauc plot of synthesized BNNS

The strong absorption peak around ~207 nm is due to the π→π* transition in BN [15,22]. It's the only peak seen in hBN, with the increment of layer number it shifts towards higher wavelength [22]. The presence of contaminants or any other chemicals during the synthesis process leads to other peaks at generally higher wavelengths. Nonetheless, the presence of only a sharp BN peak is a signature of the material synthesized is pure. The UV-Vis spectrum is used to determine the optical band gap of BNNS, from the Tauc plot. Typically Tauc plot [28] reveals the value of optical band gap in a material using the energy on the abscissa and a quantity $(ah\nu)^{1/r}$ on the ordinate, where 'a' is the absorption coefficient, and 'r' is a quantity has different values depending upon the nature of the electronic transition, here in our case we'll take r equals to ½ as for direct allowed electronic transition r=1/2. The optical band gap is then can be determined from the intercept of the X-axis (energy axis) by the tangent of the plot. Here in our case, the optical band gap is slightly greater than 5 eV. The optical band gap also reinforces the fact that the sonication reduced the layer number of BN to ~4 layers.

## 3.4. XRD studies:

The XRD plot is plotted in fig-4. The diffraction of X-ray by crystal planes of material is depicted in the XRD plot, which enables us to determine the planes in the material. Here in the XRD plot, one peak at ~26° is observed, which corresponds to the (002) plane of BNNS [29]. Due to fewer layer numbers and low dimensionality, the peak is not sufficiently sharp. It's also can be inferred that the (002) plane is abundant in the samples.

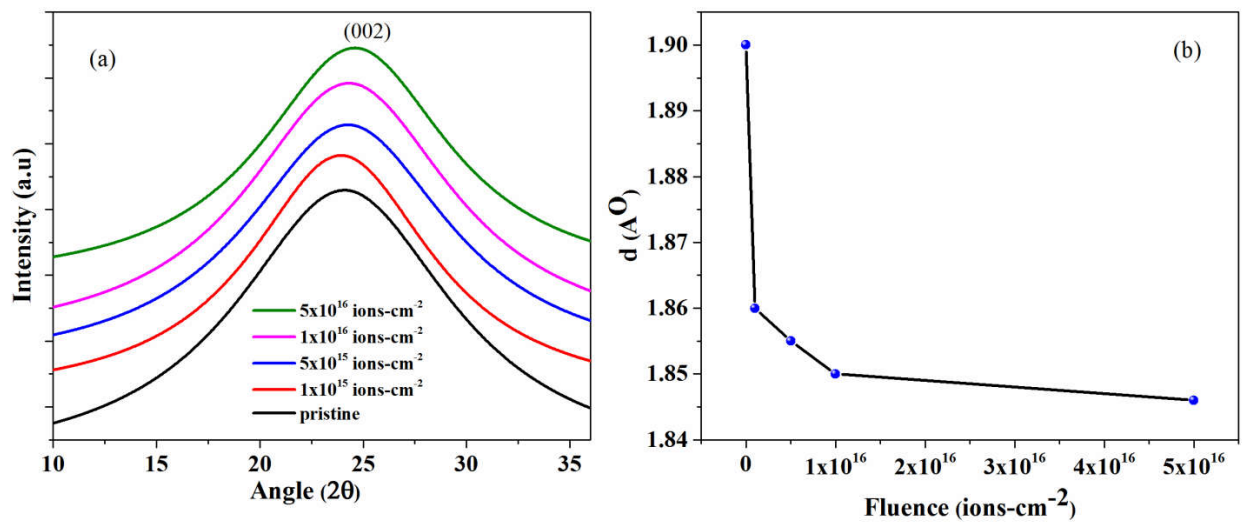

Fig-4: (a) XRD plot of pristine and ion irradiated BNNS. (b) The peak position shifting of (002) plane in BNNS after ion irradiation.

In figure-4, the XRD plot of pristine and Au irradiated BNNS is shown. The intensity was arbitrarily chosen. It's observed that the peak position is shifted towards a higher angle with increasing ion fluence. We know that for a fixed X-Ray source the inter-planer distance (d) and angle of diffraction are inversely proportional. Hence, it can be concluded that the inter-planer distance is decreased with an increasing dose of ion irradiation. The value of d of pristine and differently irradiated BNNS is given in table I.

| Fluence (ions/cm$^2$) | d (ang) | Strain (%) |
|---|---|---|
| 0 | 1.9 | ---- |
| 1E15 | 1.86 | 2.1(C) |
| 5E15 | 1.855 | 2.36 (C) |
| 1E16 | 1.850 | 2.63 (C) |
| 5E16 | 1.846 | 2.84 (C) |

Table-I: Values of inter-planer distance (d) along (002) in pristine and irradiated BNNS and corresponding strain. C stands for compressive strain.

Analyzing the FWHM of XRD plots we can also find out the strain due to ion irradiation using equation-2[2, 30].

$$\epsilon = B/4tan\theta \qquad (2)$$

Where $\epsilon$ is the strain-induced, B is the FWHM, θ is the diffraction angle.

The calculated value of strain solely due to ion irradiation is documented in table-II.

| Fluence (ions/cm$^2$) | Strain (%) |
|---|---|
| 0 | ---- |
| 1E15 | 0.8(C) |
| 5E15 | 1.55 (C) |
| 1E16 | 2.64 (C) |
| 5E16 | 2.84 (C) |

Table-II: Values of inter-planer distance (d) along (002) in pristine and irradiated BNNS and corresponding strain from FWHM of XRD. C stands for compressive strain.

It can be concluded from the values of d that, finite compressive strain along (002) direction in BNNS is induced due to Au beam irradiation.

### 3.5. Work function modulation:

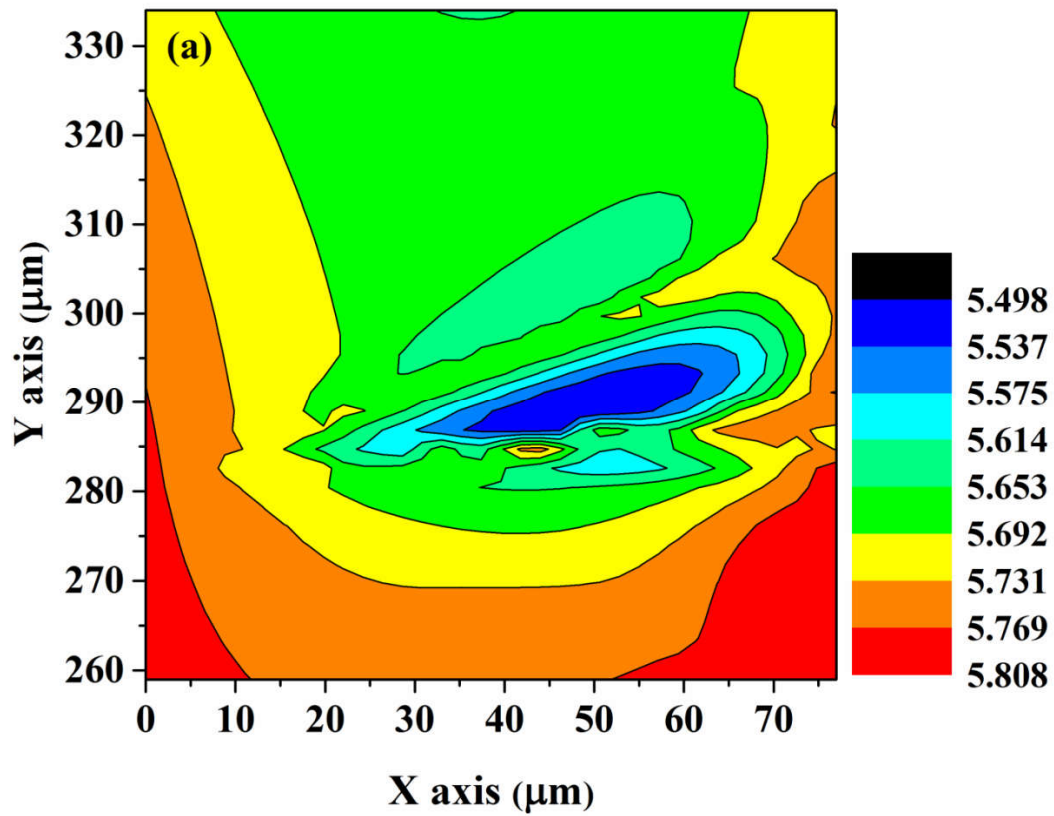

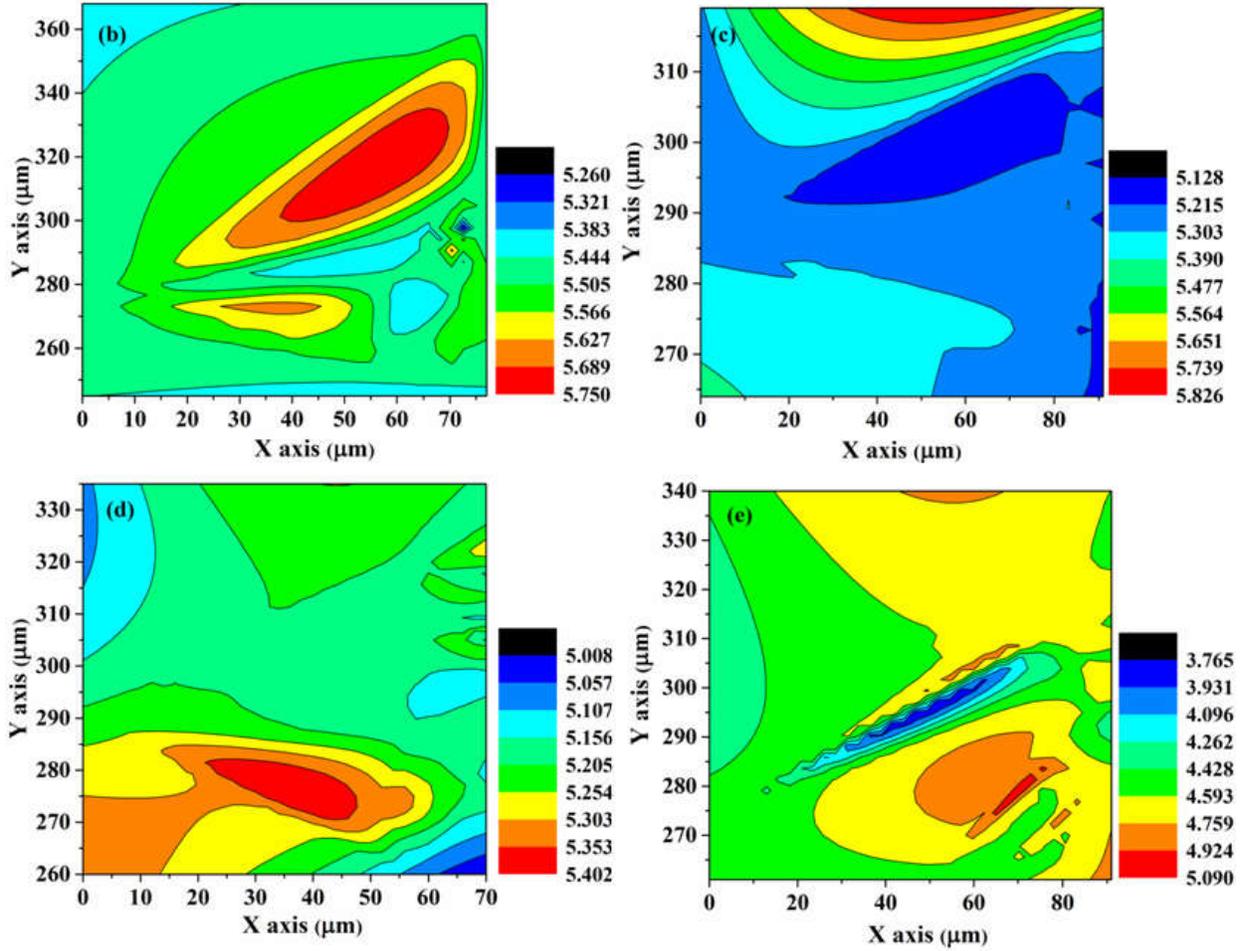

Fig-5: (a) Work function mapping of as-synthesized BNNS. (b), (c), (d), (e) Work function mapping of Au irradiated BNNS.

The work function mapping of as-synthesized BNNS is given in fig-5. It was determined using SKPM. The gold tip embedded in SKPM allows us to map the contact potential difference between the gold tip and the surface of BNNS. The relation between CPD and work function is given in equation-3[2, 20].

$$eV_{CPD} = \Phi_{tip} - \Phi_{BNNS} \qquad (3)$$

Where $\Phi_{tip,BNNS}$ is the work function value of gold tip (~5.1 eV) and BNNS. The LHS of the equation denotes the contact potential difference. The average work function of BNNS as calculated using the equation comes out to be ~5.8 eV[20, 21]. Since ion irradiation causes compressive strain along (002) direction in BNNS, it's obvious that the work function will be modified. The work function is defined as the minimum energy to excite an electron to vacuum level from Fermi level such that it'll go freely without experiencing the attraction towards the material. The work functions of differently irradiated materials are given in fig-4 (b, c, d, e). It's observed that with increasing dose the work function decreases.

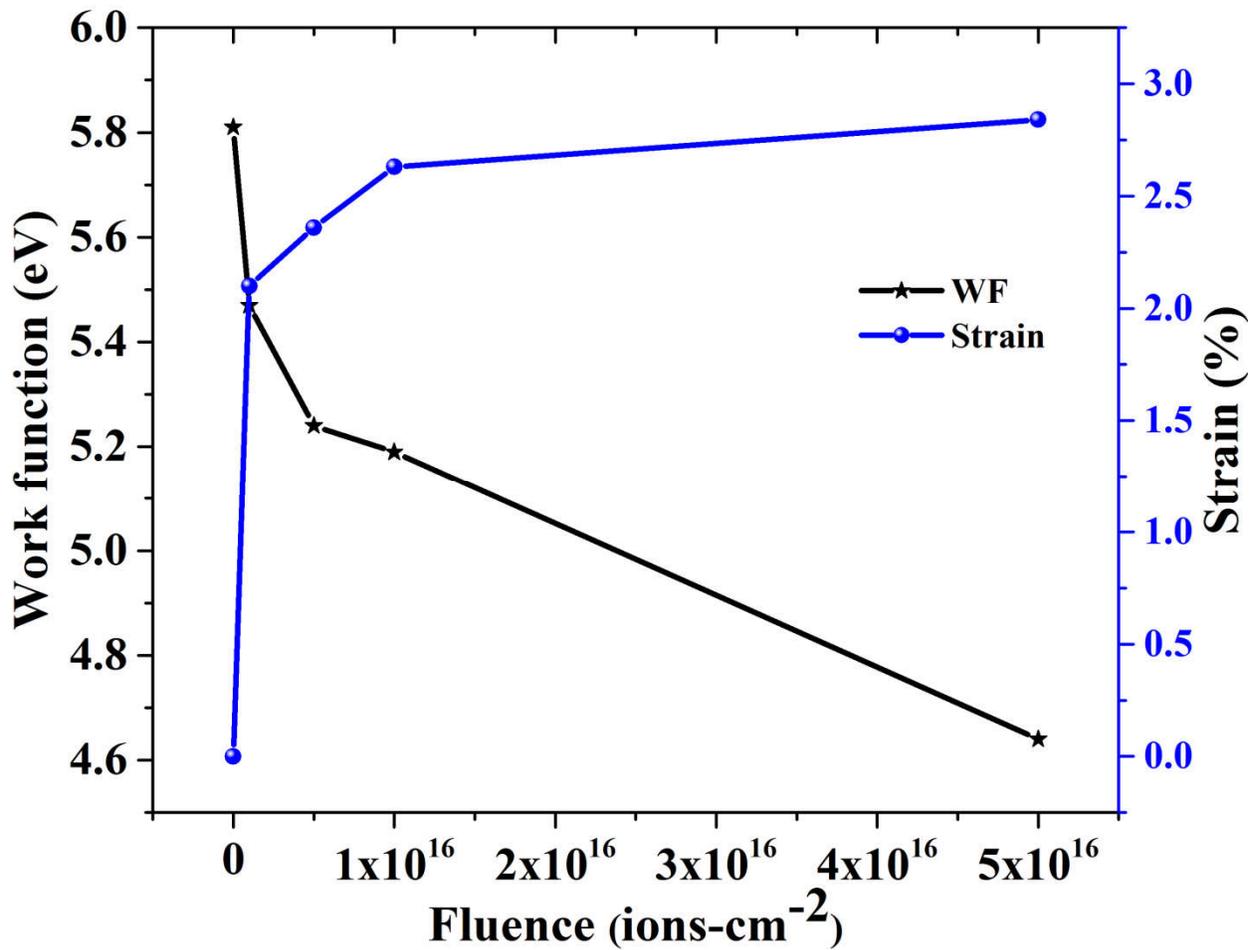

Fig-6: Modulation of work function and strain in BNNS vs ion fluence.

In the above figure (figure-6), the evolution of work function and strain with ion fluence have been depicted. The systematic compressive strain inside BNNS due to irradiation of the Au beam lowers its work function [21]. It's reported for TMDCs that compressive strain lowers the work function, whereas slight tensile strain firstly increases then decreases it [9]. In BN-MS$_2$ (M = Mo, W) compounds it's also observed that compressive strain shifts the Fermi level towards the conduction band hence lowers the work function.

## 4. Discussion:

The irradiation of Au ions (30 keV) on BNNS observed to be imposing strain in (002) direction. It is compressive as the distance between layers gets reduced. The evolution of the work function of BNNS was witnessed and certainly, it was decreased. It's thus important to investigate the interaction of low energy Au beam with BNNS and find out the explanation of the changes the BNNS went through on irradiation. The energetic beam, when exposed to the material, loses its energy in mainly two ways [31, 32, 33]. Comparatively low energetic ions mainly lose their energy to the nucleus of the target material, which is in general elastic in nature. This type of elastic collision produces recoils and point defects. If the incident ion mass and energy are high then it might produce subsequent recoils eventually lead to collision cascade. If the collision takes place near the surface then it might eject atoms from the lattice sites, known as sputtering. This type of energy loss is known as Nuclear energy loss ($S_n$). Another way of losing energy is called electronic energy loss ($S_e$), where the energy lost to target electrons however it happens at higher energies. The $S_e$, $S_n$ values of Au ion in BNNS, as well as the vacancy profile, was computed using SRIM-2013 software [34].

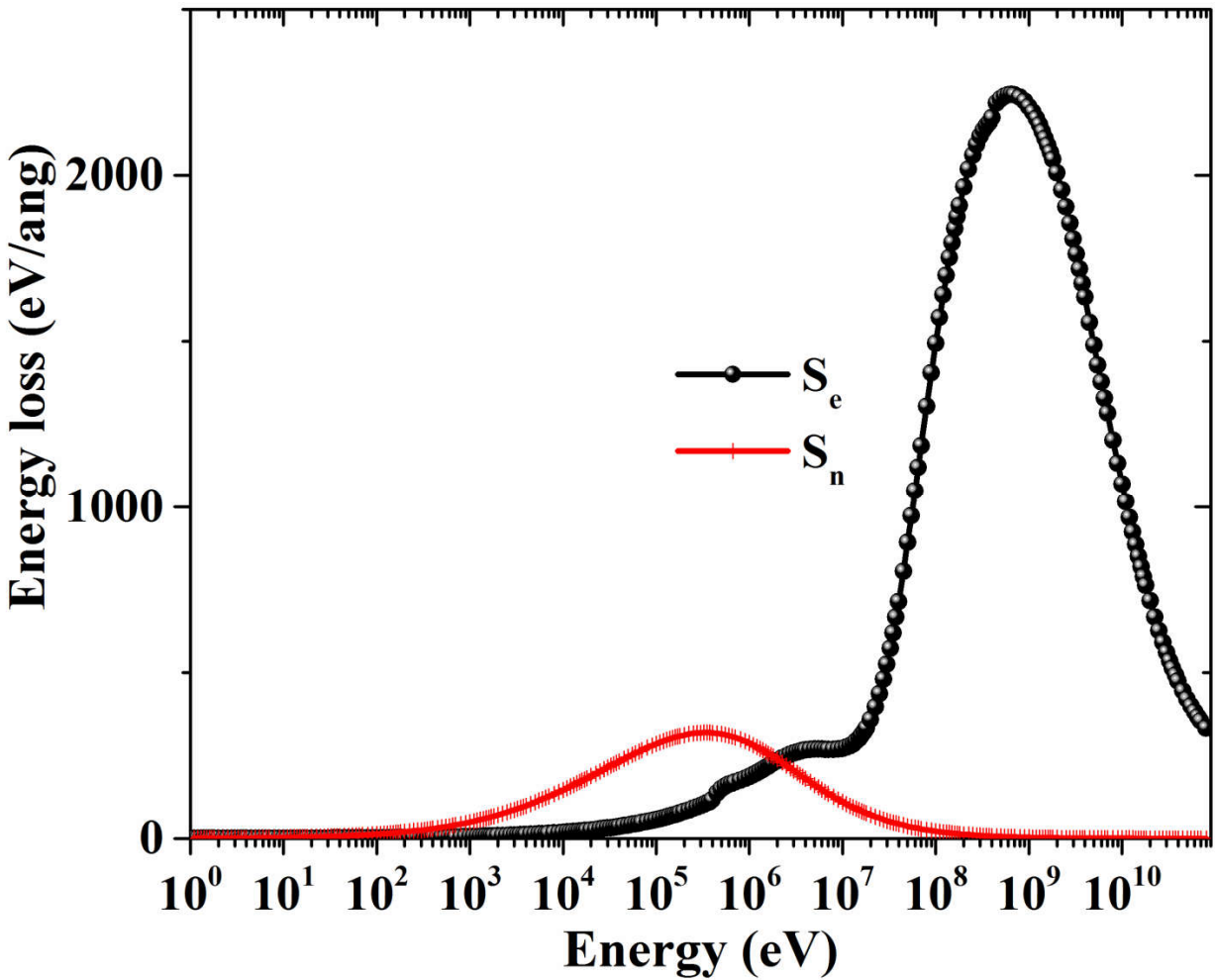

Fig-7: Electronic ($S_e$) and Nuclear energy loss ($S_n$) of 30 keV Au ions in BNNS.

The above figure (figure-7) depicts the variation of energy loss of both types of Au ion (30 keV) when irradiated with BNNS. From the figure it's observed that the nuclear energy loss is dominating in our case, hence, point defects and produce collision cascades are expected [33]. The range of the ions inside BNNS also computed, the range found was slightly greater than 200 angstrom. The thickness determined from the Raman spectrum was found to be 13.32 angstrom. Hence, mostly the surface layers of BNNS in our sample were affected by the creation of vacancies.

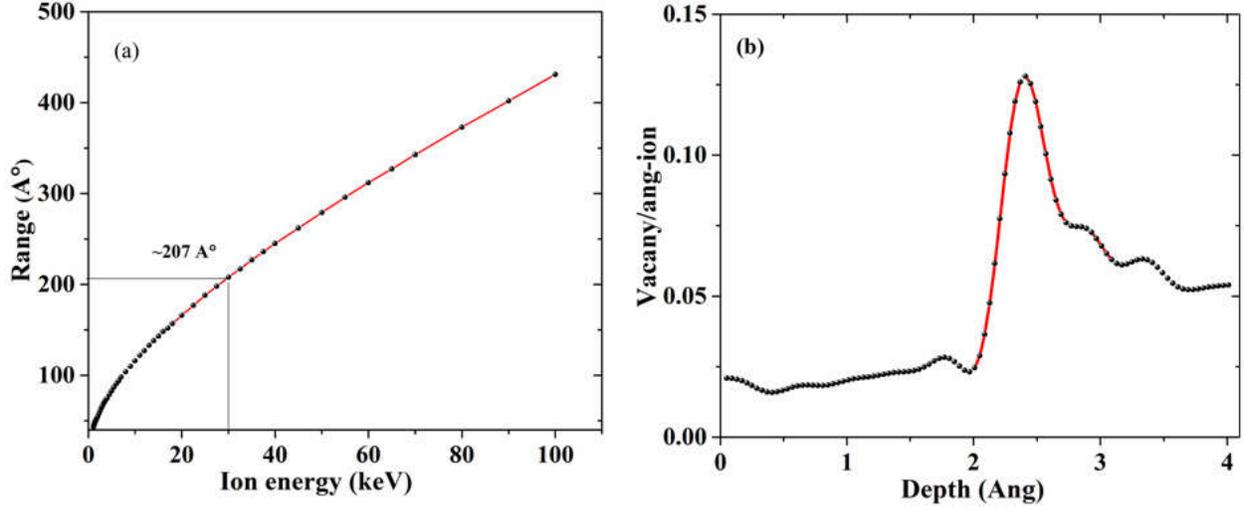

Fig-8: (a) Range of Au ion in BNNS. (b) vacancy production in BNNS by Au ions.

The surface layers of BNNS therefore modified by Au beams by the production of a few vacancies of B and N. The effect of vacancies in electronic band structure is nonetheless important yet the prompt effect is the effect of strain which BNNS went through. The appearance of strain due to vacancies of atoms was discussed in the literature [35, 36]. Briefly, the nearest neighbor distance before vacancy formation changes after the formation of vacancy. Hence, the lattice constants eventually change. In our experiment it was observed vacancy generation by Au ion irradiation lowers the lattice constant along (002) direction, which implies a compressive strain along (002) comes to play. The XRD plot supports the fact as with increasing ion fluence the peak shifts towards a higher angle. That means the inter-layer distance (d) is decreasing with increasing fluence. Thus a compressive strain is acting in (002) direction. The compressive strain modifies the electronic band structure. It was reported for TMDCs the Fermi level up shifts therefore, the work function decreases[8]. The nano-composites made of TMDCs and BN also show similar results[37]. It's probably up shifting of maximum occupying electronic bands due to the lowering of atomic distances caused by compressive strain, the effective energy to excite an electron from Fermi level to vacuum level decreases. Hence, the work function decreases. To establish the fact theoretically recall the theory of work function. The work function depends on the electrostatic potential ($V_{ref}$), vacuum level ($\varphi$) and Fermi energy ($E_f$). This can be written as [38]

$$W = -\mu + D = -(E_f - V_{ref}) - (\varphi - V_{ref}) \qquad (5)$$

Where, $E_f$ = Fermi energy, $V_{ref}$ = average electrostatic potential, $\varphi$ = vacuum level.

$$\Delta W = -\Delta\mu + \Delta D$$

$$\Delta\mu = \mu(V_s) - \mu(V_0) = E_f(V_0)\beta.$$

(Here, $V_0$ = the volume of unstrained material=$V_0(x, y, z)$; $V_s$ = the volume of strained material= $V_s(x+\lambda, y-\xi, z-\xi)$ since, $\lambda, \xi$ are small)

$$\beta = \text{relative change in atomic distance along (002)} = \frac{2}{3}\frac{\lambda}{x}.$$

$$D = -\frac{ep}{\epsilon_0} = kp = k\int_{x_0}^{L} rn(r)dr = \int_{x_0}^{L} xn(x)dx$$

Conservation of electronic charge at surface requires $n_0 V_0^{2/3} = n_s V_s^{2/3}$ which gives,

$$\Delta D = D(V_0)\beta$$

So, change in work function

$$\Delta W = \beta D(V_0) - \beta\mu(V_0) - \beta\mu(V_{ref}) = [W_0 - V_{ref}]\beta \approx \frac{2}{3}\left(\frac{\lambda}{x}\right)W_0 \qquad (6)$$

For a compressive strain, $\lambda$ is negative, which suggests that compressive strain lowers the work function.

| Fluence (ions/cm$^2$) | Strain (%) |
|---|---|
| 0 | ------ |
| 1E15 | 0.58 |
| 5E15 | 1.17 |
| 1E16 | 1.38 |
| 5E16 | 2.58 |

Table-III: Strain values from work unction values (using SKPM).

The strain values were also calculated from work function shifting of BNNS induced by Au ion beams using SKPM (table-III). The slight dissimilarity in the strain values as calculated from XRD plot and as calculated from work function shifting may be explained in terms of the slight tensile strain plays along the perpendicular plane, however small, effect of which was ignored completely. However, at higher fluence the difference in values is reduced.

## 5. Conclusion:

The synthesis of BNNS using chemical route is discussed. The synthesized BNNS was exposed to ion irradiation at low energy which subsequently creates vacancies at the surface region. The theoretical explanation of vacancy creation and induced strain along (002) direction due to atomic vacancy was discussed. The origin of compressive strain as well as its role in lowering work function is rigorously explored. It was concluded that by controlling the ion dose one can lower the work function in BNNS. The leap forward to the use of low dimensional BNNS in photoelectric cells was discussed where materials of low work function are looked for. We also have examined the possibility to compute strain in materials by analyzing the electronic work function and it comes out that qualitatively the values are reputable.